\documentclass[cits]{PoS}

\title{\boldmath Rare $B$ Decays Potential at Super$B$\unboldmath}

\ShortTitle{HQL 2010}

\author{{\bf Alexander Rakitin} \\
        California Institute of Technology\\
        E-mail: \email{arakitin@hep.caltech.edu}\\
        {\rm (on behalf of Super$B$ Collaboration)}}

\abstract{We present a short overview of the most important
          rare $B$ decay analyses which will be performed
          using $75~{\rm ab}^{-1}$ dataset which is expected to be 
          provided by Super$B$ Factory within five years from its 
          starting date.}

\FullConference{The Xth Nicola Cabibbo International Conference on Heavy Quarks and Leptons,\\
		October 11-15, 2010\\
		Frascati (Rome) Italy}

\begin{document}

\section{Motivation for Super$B$ Factory}

\subsection{Why Rare Decays?}

There are two major ways to discover physics beyond the Standard Model
(SM). The first one is to directly produce non-SM particles in
collisions and detect their decay products, the second one is to
observe the effects of these particles on the decay rates of the rare
decays of the SM bound states. The former way requires high energy,
the latter -- high statistics. The discovery of the $c$-quark proved
the validity of both ways: initially heavy $c$-quark was introduced to
bring the theoretically calculated decay rate of the rare decay of
light meson $K_L\to\mu^+\mu^-$ into agreement with the experiment~\cite{GIM}, and
only later it was found in direct production and decay of the $J/\psi$
bound state~\cite{Jpsi}. Following this historical lesson, one can look for the
effects of extremely heavy non-SM particles in the decays of much
lighter $B$-mesons.

\subsection{\boldmath Why Rare $B$ Decays?\unboldmath}

\begin{figure}[h]
\begin{center}
\includegraphics[width=0.5\textwidth]{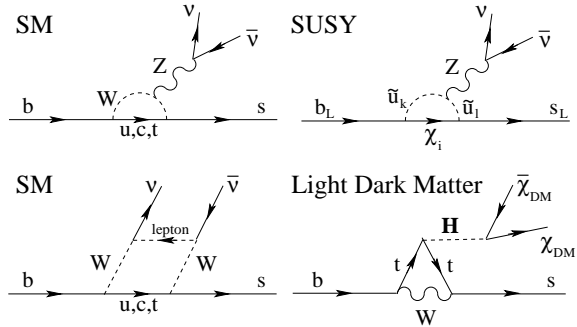}
\caption{Possible NP contributions to $b\to s\nu\bar{\nu}$ decay. }
\label{fig:diag} 
\end{center}
\end{figure}

The decays of the heavy $b$-quark involve all the lighter quarks,
therefore one has more chances to find non-SM effects. There are many
processes sensitive to these effects, including $B^0$-$\overline{B^0}$
mixing, penguin decays $b\to s\gamma$, $b\to s\ell^+\ell^-$ and $b\to
s\nu\bar{\nu}$, transitions $b\to sq\bar{q}$, $b\to dq\bar{q}$, and annihilations
$b\bar{q}\to \ell^+\ell^-$ and $b\bar{q}\to \ell\nu$. A few examples
of possible New Physics (NP) contributions to $b\to s\nu\bar{\nu}$ decay are
shown in the diagrams in Fig.~\ref{fig:diag}.

\subsection{Why Super(B)-Flavor Factory?}

The projected Super$B$ factory~\cite{SuperBdet} is a linear $e^+e^-$
collider intended to deliver instantaneous luminosity of ${\cal L} =
10^{36}~{\rm cm}^{-2}~{\rm s}^{-1}$ at the same center-of-mass energy
as the current $B$-factories: the mass of $\Upsilon(4S)$. In five
years ($1.5\times10^{8}~{\rm s}$) of running with estimated
$\sim\!50\%$ downtime Super$B$ is expected to collect $\sim\!75~{\rm
  ab}^{-1}$ of data.  This is two orders of magnitude larger than the
currently available datasets collected by BABAR ($0.53~{\rm ab}^{-1}$)
and Belle ($0.95~{\rm ab}^{-1}$). This humongous dataset will
drastically improve our chances to find the New Physics effects in
rare $B$ decays\footnote{as well as rare $D$ and $\tau$ decays -- for
  those see the corresponding peer proceedings.}. And if (when) LHC
finds the New Physics before the start of Super$B$, the flavor content
of this New Physics will still have to determined, and Super$B$ will
be playing major role in this.

\section{Experimental Technique}

\begin{figure}[h]
\begin{center}
\includegraphics[width=0.4\textwidth]{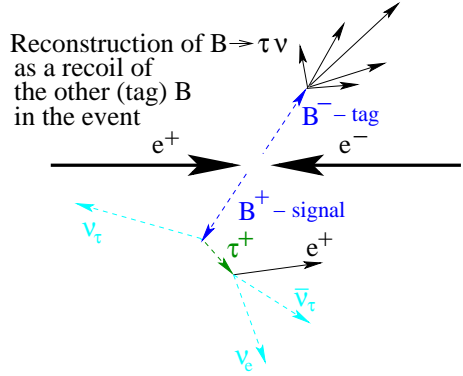}
\caption{Rare $B$ decays with one or more neutrinos in the final state can only be done in $e^+e^-$ machines.}
\label{fig:B-to-tau-nu} 
\end{center}
\end{figure}

Fully-reconstructible rare $B$ decays (such as $B_{(s)}\to\mu\mu$ or
$B_{(s)}\to K^{(*)}\mu\mu$) can be analysed at both hadronic machines
(LHC, Tevatron) and Super$B$. But those rare $B$ decays which contain
one or more neutrinos in the final state can only be investigated in
clean $e^+e^-$ environment. The corresponding data analyses at Super$B$
will be exploiting the same recoil technique which has been
extensively used in $B$-factories: full or partial reconstruction of
one $B$-meson (the {\it tag} $B$) and ascribing all the remaining
objects to the second $B$-meson (the {\it signal} $B$) -- see
Fig.~\ref{fig:B-to-tau-nu}.  

The tag $B$ meson can be reconstructed in
two ways. In the first one it is fully-reconstructed in the decay into
pure hadronic final state.  This technique gives better kinematic
constraint of the $B$ meson but has a lower reconstruction efficiency
($\sim\!0.4\%$).  In the second way, the tag $B$ meson is partially
reconstructed in the decay into semileptonic final state with missing
neutrino. Obviously, in this case the kinematic constraint is much
worse, but the reconstruction efficiency is higher ($\sim\!2.0\%$).

A very important quantity here is so-called $E_{extra}$ -- the sum of
all the energy depositions in the Electromagnetic Calorimeter (EMC)
not associated with any physics objects. In a well-reconstructed event
$E_{extra}$ must be of the order of 200-300~${\rm MeV}/c^{2}$
(calorimeter noise), but in the events where we actually miss a
particle it can be larger. Super$B$ will have an additional backward
end-cap EMC with respect to BABAR EMC configuration, to diminish the
amount of lost particles.

\section{Rare $B$ Decays}

\subsection{$b\to s\nu\bar{\nu}$}

Theoretically, it is better to measure inclusive decay rate
$B\to X_{s}\nu\bar{\nu}$, but such analysis would be challenging from
experimental point of view. For this reason we are going to measure
the rates of the exclusive decays $B_u\to K\nu\bar{\nu}$ and $B_d\to
K^{*0}\nu\bar{\nu}$. The latter decays has another observable --
longitudinal polarization $\langle F_{L}\rangle$ properly averaged
over the neutrinos' invariant mass~\cite{Altmannshofer:2009ma} -- which
is easy to calculate. The current limits of the decay rates are about
an order of magnitude larger than SM predictions (Tab.~\ref{tab:svv}).

\begin{table}[h]
\begin{center}
\begin{tabular}{lll}
\hline
Observable  & SM prediction &  Experiment \\
\hline
$\mathcal B (B^0 \to K^{*0} \nu\bar\nu)$ & $( 6.8^{+1.0}_{-1.1} ) \times 10^{-6}$~\cite{Altmannshofer:2009ma} & $< 80 \times 10^{-6}$ \cite{:2008fr} \\
$\mathcal B(B^+ \to K^+ \nu\bar\nu)$   & $( 3.6 \pm 0.5 ) \times 10^{-6}$~\cite{Bartsch:2009qp} & $< 14 \times 10^{-6}$ \cite{:2007zk} \\
$\mathcal B(\bar B \to X_s \nu\bar\nu)$ & $( 2.7\pm0.2 ) \times 10^{-5}$ \cite{Altmannshofer:2009ma} & $< 64 \times 10^{-5}$ \cite{Barate:2000rc} \\
$\langle F_L(B^0 \to K^{*0} \nu\bar\nu) \rangle$ & $0.54 \pm 0.01$ \cite{Altmannshofer:2009ma} & --  \\
\hline
\end{tabular}
\caption{SM predictions and experimental 90\% C.L. upper bounds for the four $b\to s\nu\bar{\nu}$ observables.}
\label{tab:svv}
\end{center}
\end{table}

\begin{figure}[h]
\centering
\includegraphics[width=0.45\textwidth]{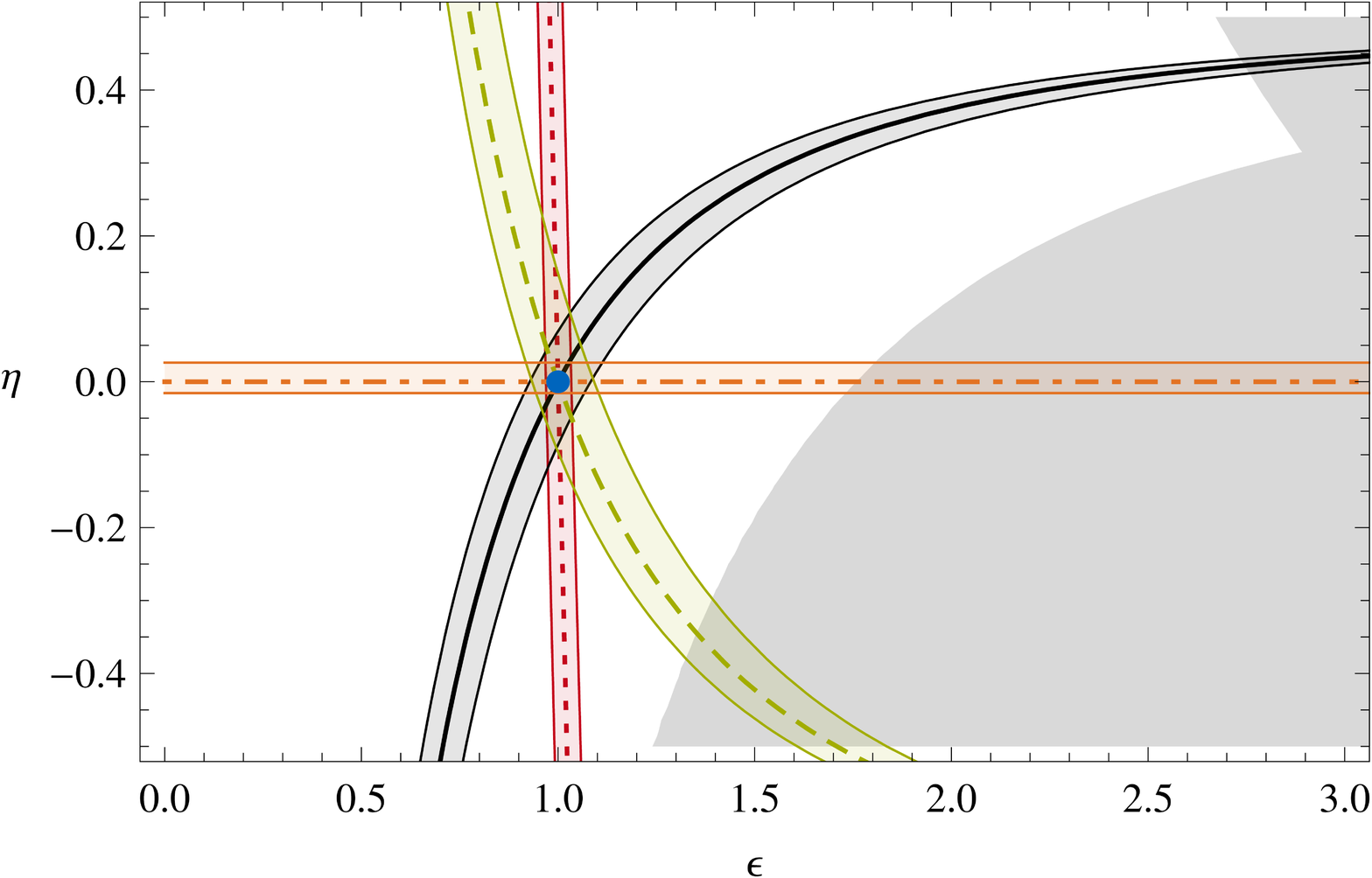}
\includegraphics[width=0.45\textwidth]{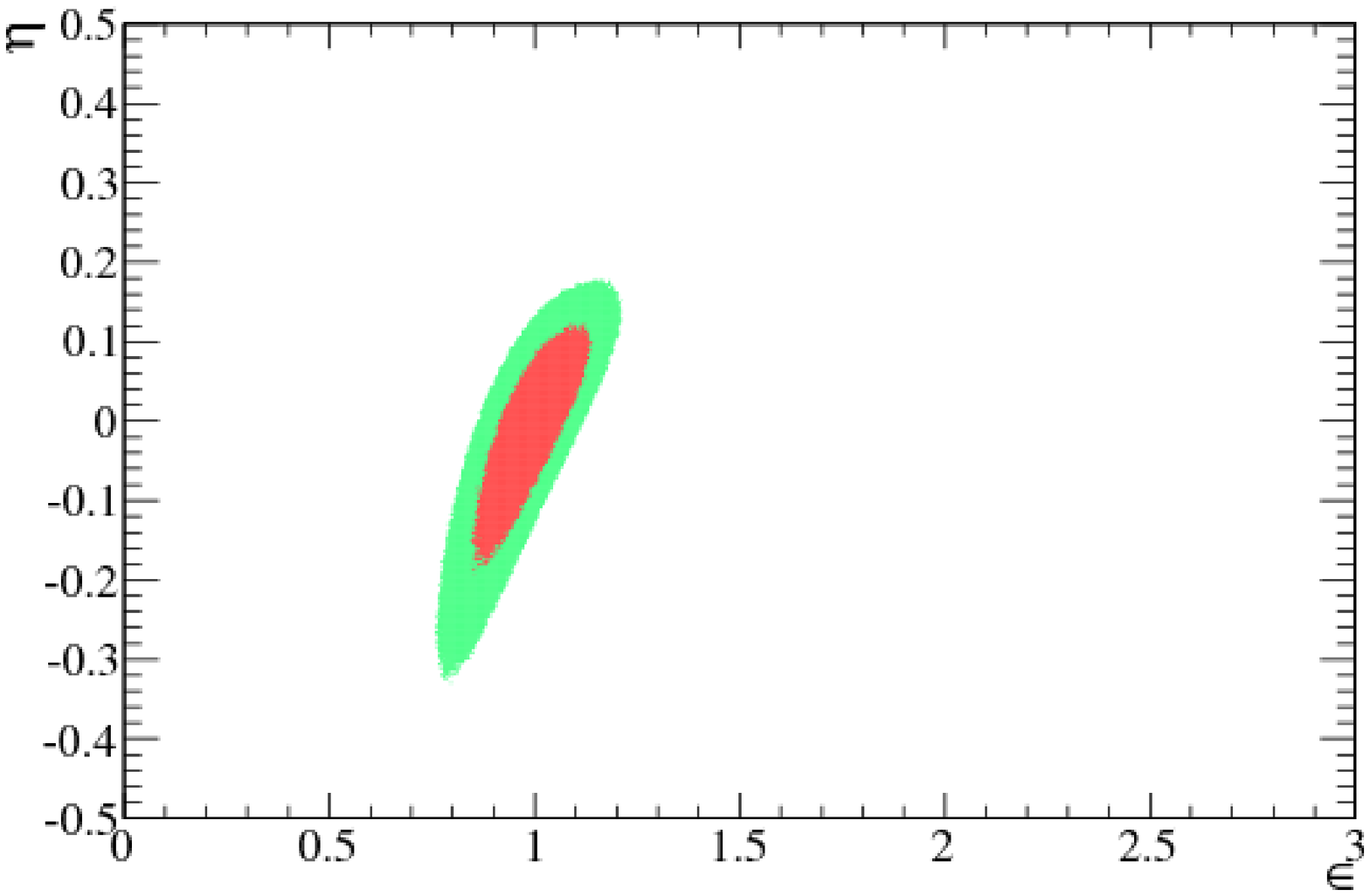}
\caption{{\bf Left:} Theoretical (4 colored bands) and 90\% C.L. 
experimental (grey area) constraints on $\epsilon$ and $\eta$.
The green band (dashed line) shows the constraint from $\mathcal B(B \to K^* \nu\bar\nu)$,
the black band (solid line) -- $\mathcal B(B \to K \nu\bar\nu)$,
the red band (dotted line) -- $\mathcal B(B \to X_s \nu\bar\nu)$,
and the orange band (dot-dashed line) -- $\langle F_L \rangle$. 
{\bf Right:} Expected 68\% C.L. (green) and 95\% C.L. (red) constraints on $\epsilon$ and $\eta$ 
from $75~{\rm ab}^{-1}$ at Super$B$.}
\label{fig:epseta}
\end{figure}

It is possible to parametrize the deviations of all the branching
fractions from their SM values in terms of only two phenomenological
parameters $\epsilon$ and $\eta$ which are respectively equal to 1 and 0 in SM:
\begin{eqnarray}
\label{eq:epseta-BKsnn}
 R(B \to K^* \nu\bar\nu) & =&  (1 + 1.31 \,\eta)\epsilon^2, \\
\label{eq:epseta-BKnn}
 R(B \to K \nu\bar\nu)   & =& (1 - 2\,\eta)\epsilon^2, \\
\label{eq:epseta-BXsnn}
 R(\bar B \to X_s \nu\bar\nu) & =& (1 + 0.09 \,\eta)\epsilon^2, \\
\label{eq:epseta-FL}
 \langle F_L \rangle/\langle F_L \rangle_{\rm SM}             & =&  \frac{(1 + 2 \,\eta)}{(1 + 1.31 \,\eta)},
\end{eqnarray}
where $R(X)=\mathcal B(X)/\mathcal B(X)_{\rm SM}$. Notice that
$\langle F_L \rangle$ does not depend on $\epsilon$. An experimental
observation of such a dependence would be a sign for the existence of
the right-handed currents. The plots in Fig.~\ref{fig:epseta}
(Refs.~\cite{SuperB,ABevan}) demonstrate current theoretical and experimental
constraints in $\epsilon$-$\eta$ plane.

\subsection{$B\to\ell\nu$}

\begin{figure}[h]
\centering
    \includegraphics[width=4.cm,angle=-90]{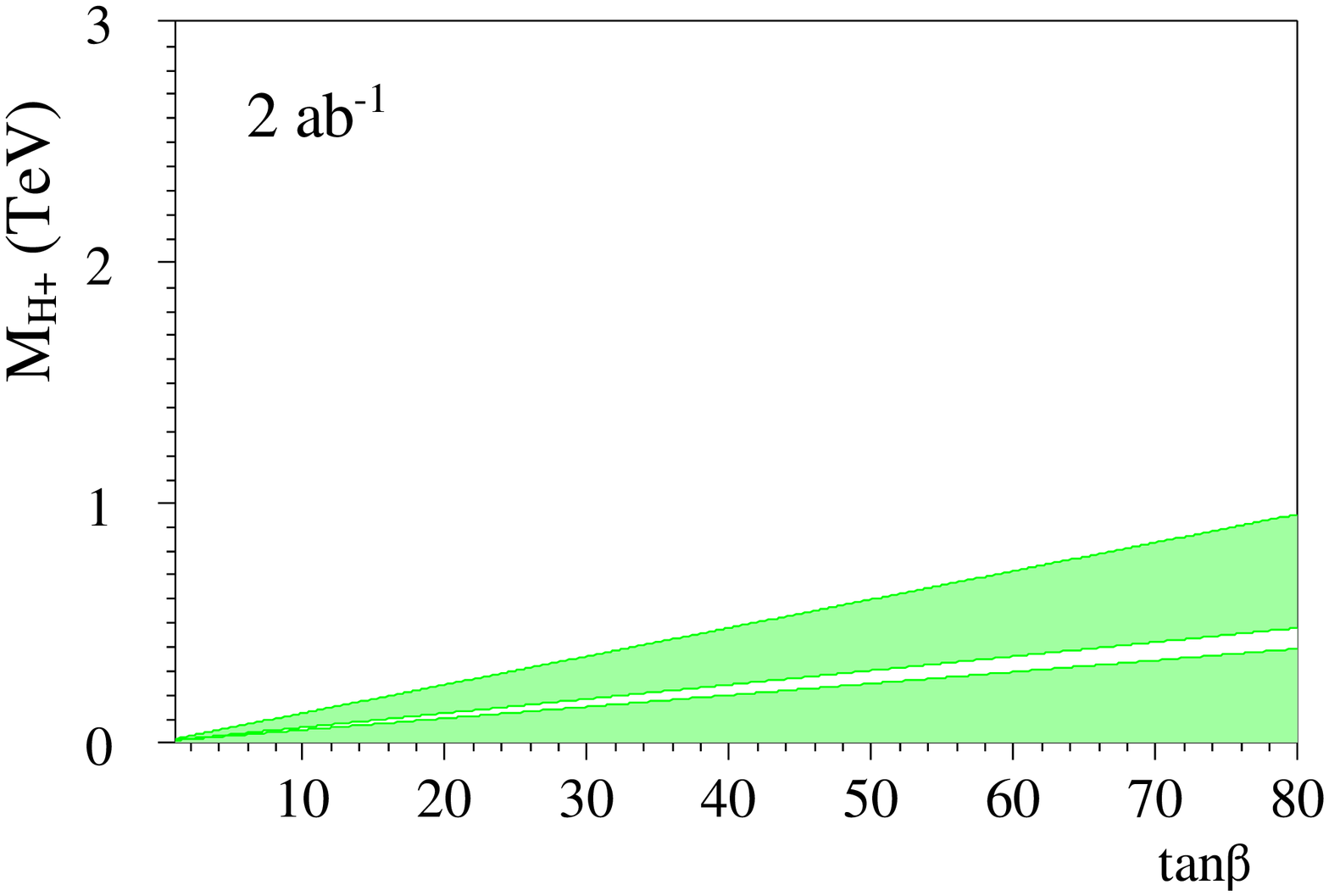}
    \includegraphics[width=4.cm,angle=-90]{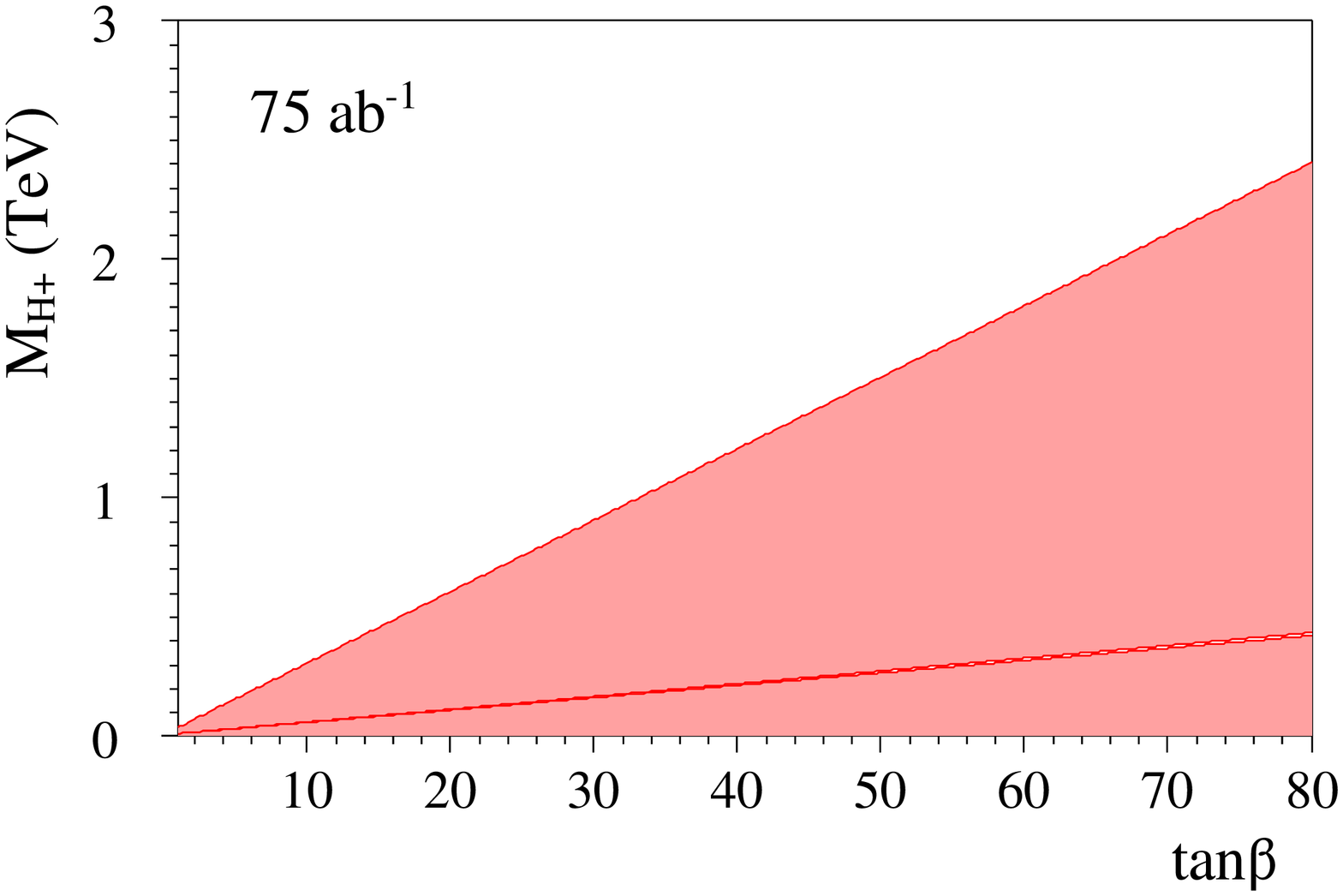}
    \caption{Excluded regions for the currently available total dataset
  ($2~{\rm ab}^{-1}$) and for the expected dataset from Super$B$
  ($75~{\rm ab}^{-1}$) for both $B\to\tau\nu_\tau$ and
  $B\to\mu\nu_\mu$ analyses together.}
\label{fig:2and75}
\end{figure}

This annihilation proceeds via charged current and is sensitive to the charged
Higgs, especially for the large values of $\tan\beta$. The Two Higgs Double Model (2HDM) yields
$$
\frac{\mathcal B(B \to \ell\nu_\ell)_{2HDM}}{\mathcal B(B \to \ell\nu_\ell)_{SM}}
= \left(1 - \frac{m_B^2\tan^2\beta}{m^2_{H^\pm}}\right)^2.
$$
The $\mathcal B(B\to\tau\nu_\tau)$ is within the reach of current
$B$-factories and has already been measured by BABAR~\cite{BABAR} and
Belle~\cite{Belle} 
but Super$B$ is expected to significantly improve the precision, as well
as measure the $\mathcal B(B\to\mu\nu_\mu)$. The plots in
Fig.~\ref{fig:2and75} demonstrate the excluded regions for $\tan\beta$
and the mass of the charged Higgs for the currently available total
dataset ($2~{\rm ab}^{-1}$) and for the expected dataset from Super$B$
($75~{\rm ab}^{-1}$) for both $B\to\tau\nu_\tau$ and $B\to\mu\nu_\mu$
analyses together. With further increase of the dataset beyond
$75~{\rm ab}^{-1}$ the uncertainty on $\mathcal B(B\to\tau\nu_\tau)$
becomes systematics-dominated while $\mathcal B(B\to\mu\nu_\mu)$ keeps
scaling with statistics.

\vspace*{0.3cm}

%

\subsection{$B\to\ell\nu\gamma, B_{(s)}\to\ell\ell\gamma, B_{(s)}\to\gamma\gamma$}

The branching fractions of these radiative decays have no hadronic
uncertainties and no helicity suppression (due to the presence of the
photon).  The SM predictions together with current experimental limits
on the branching fractions are given in Tab.~\ref{tab:xxx}. Super$B$
will be able to improve the existing measurements on $B_{(s)}$ decays
and, possibly, setup a limit on $B_s\to \ell\ell\gamma$ decay.

\begin{table}
\centering
\begin{tabular}{lcc}
\hline
Process & SM Branching Fraction & Experiment 90\% C.L. \\
\hline
$B \to \ell \nu \gamma$ & $\mathcal O(10^{-6})$~\cite{valencia} & $1.56\times10^{-5}$~\cite{PDG} \\
$B \to \ell \ell \gamma$ & $\mathcal O(10^{-10})$~\cite{valencia} & $\sim\!10^{-7}$~\cite{PDG} \\
$B_{s} \to \ell \ell \gamma$ & $\mathcal O(10^{-9})$~\cite{valencia} & -- \\
$B \to \gamma \gamma$ & $\mathcal O(10^{-8})$~\cite{valencia} & $6.2\times10^{-7}$~\cite{PDG} \\
$B_s \to \gamma \gamma$ & $(2$-$8)\times 10^{-7}$~\cite{Reina:1997my} & $8.7\times 10^{-6}$~\cite{PDG} \\
\hline
\end{tabular}
\caption{SM predictions and current experimental limits on the radiative
branching fractions.}
\label{tab:xxx}
\end{table}

\subsection{$b\to s\gamma$}

\begin{figure}[h]
  \begin{center}
    \includegraphics[width=0.45\textwidth]{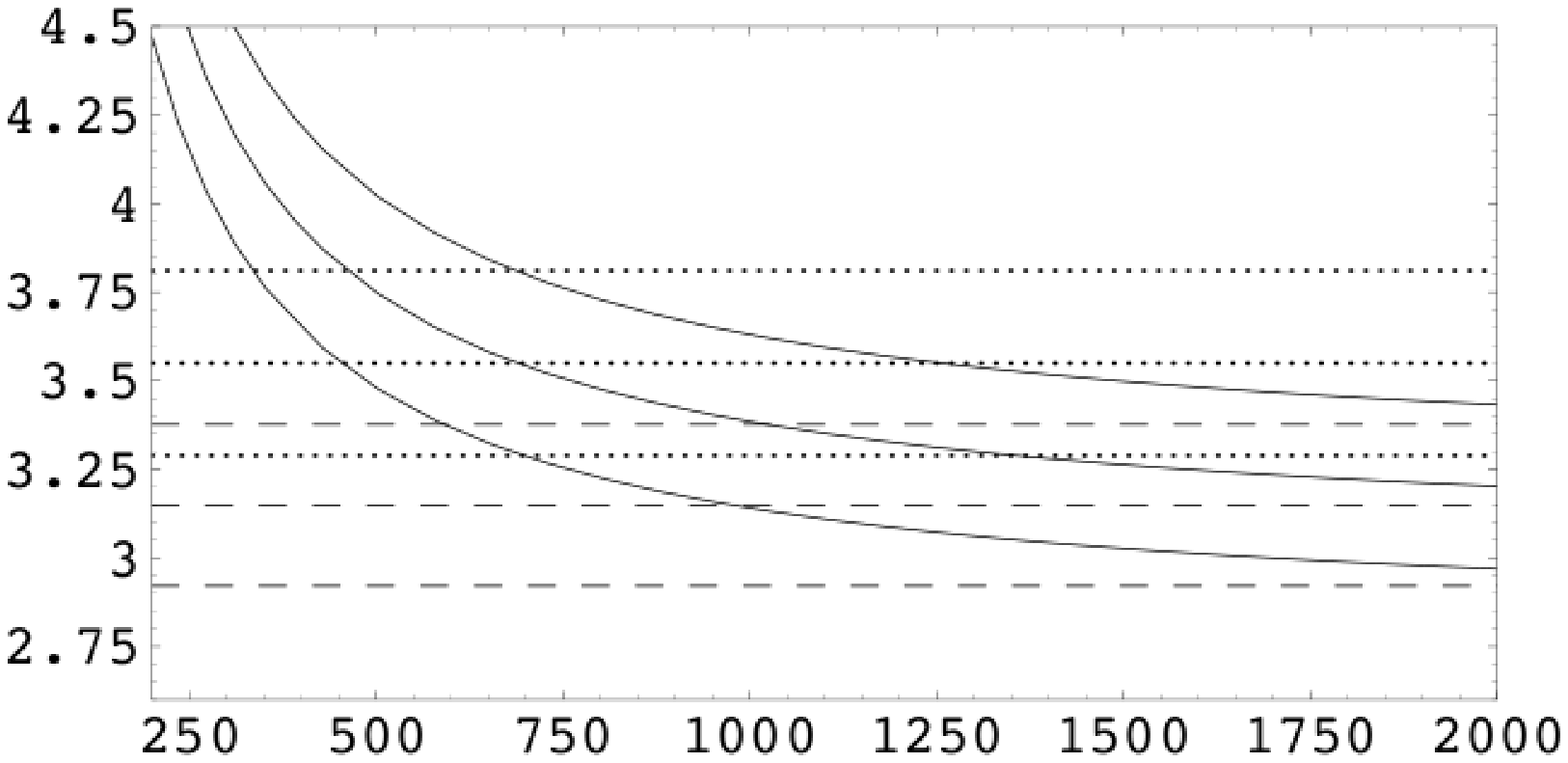}
    \includegraphics[width=0.35\textwidth]{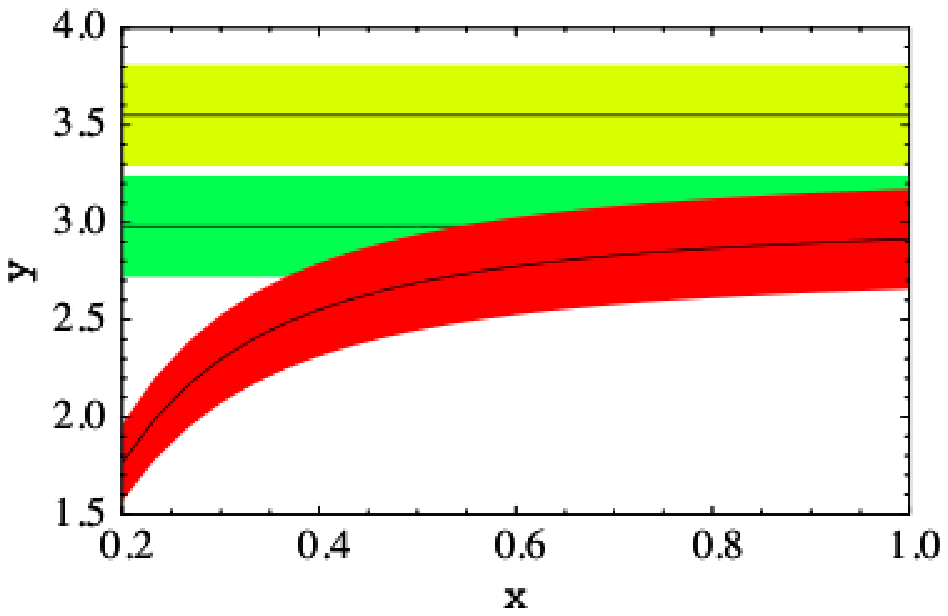}
  \end{center}
  \caption{
    {\bf Left:} ${\cal B}(\bar B \to X_s \gamma)~[10^{-4}]$ as a function of the charged
    Higgs boson mass $M_{H^{+}}~[GeV/c^{2}]$ in the 2HDM~II for $\tan\beta=2$ (solid lines).
    The dashed lines represent SM (central value plus-minus uncertainty), the dotted lines -- experiment.
    {\bf Right:}
    $y \equiv {\mathcal B} (\bar{B} \to X_s \gamma)~[10^{-4}]$ as a 
    function of inverse compactification radius $x \equiv 1/R~[{\rm TeV}]$ of mUED.
    The red (dark gray) band corresponds to the LO mUED result.
    The yellow (light gray) band -- 68\% C.L. experimental range,
    the green (medium gray) band -- SM.  
  }
  \label{fig:MHc}
\end{figure}

Similar to $b\to s\nu$, this decay is sensitive to charged Higgs in
2HDM-II model. The bound on the charged Higgs mass from the left plot
in Fig.~\ref{fig:MHc} ($M_{H^{+}} = 295~{\rm GeV}/c^{2}$ at
$95\%~C.L.$) is the strongest available
one~\cite{Misiak:2006zs}. Another application of $b\to s\gamma$ decay
-- extra dimensions. The bound on the inverse compactification radius
$1/R$ at $95\%$ C.L. in the minimal Universal Extra
Dimension model (mUED) shown in the right plot Fig.~\ref{fig:MHc} from
Ref.~\cite{Haisch:2007vb}. Super$B$ will definitely improve both plots
with the expected systematic uncertainty (3\%) dominating over the
statistical one. It is also worth mentioning here that in producing
these plots the $B$-factories used only more efficient semileptonic
tag $B$, while Super$B$ will be able to use more
kinematically-constrained hadronic tag $B$ as well.

%
%
%

\subsection{$b\to s\ell\ell$}

This rare decay is sensitive to the right-handed currents and
multitude of different types of NP. There are two approaches to this decay:
reconstruction of all the exclusive decay modes and inclusive recoil
analysis with two leptons in the recoil.  The first method is quite
complicated experimentally and not so clean from the theoretical point
of view.  The second method is better theoretically, but has a
disadvantage of low $B$-tagging efficiency and admixture of $b\to d\ell\ell$. 
Therefore, the third approach is usually used -- exclusive reconstruction
of only two modes: $B_u\to K\ell\ell$ and $B_d\to K^{*0}\ell\ell$. 
By simple extrapolation we expect an order of magnitude better
statistical uncertainty and a factor of two better systematic
uncertainty in $75~{\rm ab}^{-1}$ data sample in Super$B$ than
in the corresponding BABAR analyses.

\subsection{\boldmath Super$B$ Performance at $75~{\rm ab}^{-1}$\unboldmath}

Finally, the uncertainties on the branching fractions of the most
important rare $B$ decays projected to $75~{\rm ab}^{-1}$ dataset
expected at Super$B$ are presented in Tab.~\ref{tab:final} (Refs.~\cite{MRama,cdr,Browder:2007gg,Browder:2008em}). As we can
see, a significant improvement or a completely new measurement is
expected for all the channels.

\begin{table}[h]
\centering
\begin{tabular}{lcc}
\hline
Mode & \multicolumn{2}{c}{Sensitivity} \\
   & Current & Expected ($75 \ {\rm ab}^{-1}$)  \\
\hline
 ${\cal B}(B \to X_s \gamma)$ & 7\% & 3\% \\
 ${\cal B}(B^+ \to \tau^+ \nu)$ & 30\% & 3--4\% \\
 ${\cal B}(B^+ \to \mu^+ \nu)$ & not measured & 5--6\% \\
 ${\cal B}(B \to X_s \ell^+\ell^-)$ & 23\% & 4--6\% \\
 ${\cal B}(B \to K \nu \overline{\nu})$ & not measured & 16--20\% \\
 \hline
\end{tabular}
\caption{The uncertainties on the branching fractions of the most important
  rare $B$ decays projected to $75~{\rm ab}^{-1}$ dataset expected at Super$B$.}
\label{tab:final}
\end{table}

\section{Summary}

In conclusion, the Super$B$ Factory will be able to either spot New
Physics in the rare $B$ decays or to investigate flavor content of the
New Physics found at LHC, within only a few years after its start!

\section{Acknowledgments}

The author would like to thank the conference organizers for the opportunity to present this poster.

\end{document}